# Use of the Meta-analysis and Kolmogorov criteria in the Finding of Singularities of a Nuclear Matter Created in Ultra-relativistic Nuclear Collisions

V. A. Kizka

Department for Nuclear Power Plants Decommissioning, Institute for Safety Problems of Nuclear Power Plants, Chornobyl, Ukraine

Department of Experimental Nuclear Physics, V.N.Karazin Kharkiv National University, Kharkiv, Ukraine

*Email:* *v.kizka@ispnpp.kiev.ua*, *valeriy.kizka@karazin.ua*

**Abstract:** Published theoretical data from several models – PHSD/HSD both with and without chiral symmetry restoration (CSR), applied to experimental data on nuclear collisions from BEVALAC/SIS to LHC energies were analyzed using meta-analysis and Kolmogorov criteria. This made it possible to localize possible features of nuclear matter created in central nucleus-nucleus collisions. Ignition of a drop of quark-gluon plasma (QGP) begins already at an energy of about $\sqrt{s_{NN}}$ = 2 GeV. We estimate that this QGP droplet occupies a small fraction, 15% (average radius of about 5.3 fm, if the fireball radius is 10 fm), of the total volume of the fireball created at $\sqrt{s_{NN}}$ = 2.7 GeV. A drop of exotic matter undergoes a split phase transition – separated boundaries of sharp (1$^{st}$ order) crossover and CSR in chiral limit, between QGP and Quarkyonic matter at an energy about $\sqrt{s_{NN}}$ = 3.5 GeV. The critical endpoint of 2$^{nd}$ order probably cannot be reached in nuclear collisions. The triple phase area occupies interval from $\sqrt{s_{NN}}$ =12 GeV to 15 GeV, the critical endpoint of 1$^{st}$ order – at around $\sqrt{s_{NN}}$ = 20 GeV. The boundary of smooth (2$^{nd}$ order) crossover transition with CSR in chiral limit between Quarkyonic matter and QGP was localized between $\sqrt{s_{NN}}$ = 9.3 GeV and 12 GeV, and between Hadronic and QGP in the interval from $\sqrt{s_{NN}}$ = 15 GeV to 20 GeV, the boundary of sharp (1$^{st}$ order) crossover transition with CSR in chiral limit between Hadronic matter and QGP was localized after $\sqrt{s_{NN}}$ = 20 GeV. The phase trajectory of the hadronic corona, enveloping the exotic droplet, always remains in the hadronic phase. The possible phase diagram of nuclear matter created in mid-central heavy ion collisions is also presented in the same energy range as for central collisions. Taking into account the quantum nature of the fireball created in nuclear collisions, we also emphasize on the existence of events in central nuclear collisions at energy range from $\sqrt{s_{NN}}$ = 2 GeV to 2.76 TeV at which no exotic matter is created and nuclear matter in the fireball remains in the hadronic phase throughout its (fireball) evolution.

**Keywords:** Quark-Gluon Plasma, Quarkyonic matter, heavy ion collisions, QCD phase diagram, Kolmogorov criteria.

## 1. Introduction

Meta-analysis (analysis of analyses) was widely used already in the 18th and 19th centuries by Laplace [1] and astronomers [2]. This idea arose among astronomers even earlier, in the 17$^{th}$ century, and was further developed, especially after Blaise Pascal's invention of mathematical methods used for gambling.

The main goal of this work is to elucidate the possible phase diagram of strongly interacting matter, using existing published material obtained over several decades. To do this, we collected data for a large set of physical observables measured in relativistic nuclear collision experiments. By grouping these data for each type of particles, we calculated Kolmogorov criteria for the agreement of various theories with experiment. Averaging these criteria over all physical observables for all types of particles at each energy of nuclear collisions made it possible to construct graphs of the dependence of the averaged Kolmogorov criteria on the energy of nuclear collisions. The next step is to map these graphs with theoretical phenomenologies. This work is a continuation of similar works performed using graphs of averaged chi-square criteria [3] and averaged chi-square and relative criteria [4].

In addition, we took into account the quantum nature of the fireball resulting from relativistic nuclear collisions, discussed in [4], [5]. Accordingly, our results correspond to only one scenario of fireball evolution, which occurs with a certain probability. The probability, for example, of ignition of QGP and the type of transition of hadronic matter to deconfined one depends on the energy of nuclear collisions and the centrality of collisions. Thus, even at RHIC energy $\sqrt{s_{NN}}$ = 63 GeV, there is a non-zero probability of the formation of a fireball in central collisions of heavy ions without ignition of the QGP phase [4]. The number of these events is small at this energy, but their number increases as the energy of the central collisions decreases to the SPS energies. At the lowest RHIC energy $\sqrt{s_{NN}}$ = 3 GeV, where we expect phase transitions in nuclear matter, at least in central heavy ion collisions [6], the number of events with and without the formation of exotic matter may be comparable. This makes it impossible to adequately describe the experimental data. Theoretical models that contradict each other may all be valid, since the physical processes represented by these models are realized in nuclear collisions, but with different probabilities.

## 2. Justification of the method

Let us consider some phenomenon *P* of arbitrary nature. Let the observation of the phenomenon *P* be carried out through measurements of some set of physical observables: $S_B = \{B_1, ..., B_x\}$. Let the experiment be able to measure any observable $B_i \in S_B$ in certain interval of (kinematical) parameters determined by the experimental conditions, and this interval is divided into $n_{B_i}$ bins (the number of data points) with a widths determined by the sensitivity of experiment to $B_i$. The set of experimentally measured points of $B_i$ is $B_i^{\exp} = \{B_{i,1}^{\exp} \pm \sigma^{\exp}(B_{i,1}), ..., B_{i,n_{B_i}}^{\exp} \pm \sigma^{\exp}(B_{i,n_{B_i}})\}$, where $\sigma^{\exp}(B_{i,j})$ is the experimental error corresponding to the $j^{th}$ data point of $B_i$.

Let there be a set of theories $S_T = \{T_1, ..., T_k\}$ describing the phenomenon *P* under consideration. Let the theory $T_i$ allows to calculate all set of the observables $S_B$. For each $B_j^{\exp}$, a theoretical value should be calculated $B_j(T_i) = \{B_{j,1}(T_i) \pm \sigma(B_{j,1}(T_i)), ..., B_{j,n_{B_i}}(T_i) \pm \sigma(B_{j,n_{B_i}}(T_i))\}$, $\forall T_i$, where $\sigma(B_{j,f}(T_i))$ is the theoretical error corresponding to the $f^{th}$ data point of $B_{j,f}(T_i)$. Let us find for all $B_j^{\exp}$ and $B_j(T_i)$ the maximal deviation/distance *F* (in the sense of the statistical distance used in the Kolmogorov (or Kolmogorov-Smirnov) criterion [7]) between their data points:

$$F(B_j, T_i) = \sup \left| B_{j,k}^{\exp} - B_{j,k}(T_i) \right| \equiv \max \left| B_{j,k}^{\exp} - B_{j,k}(T_i) \right|. \qquad (1)$$

In Figure *1* we explain the calculation of the Kolmogorov criterion (1), where $B_j \equiv dN(\bar{\Lambda})/dy$, $T_i$ is some theory *T*, $B_{j,k}^{\exp} \equiv \left.\frac{dN^{\exp}(\bar{\Lambda})}{dy}\right|_{y=-0.4}$, $B_{j,k}(T_i) \equiv \left.\frac{dN^{theory}(\bar{\Lambda})}{dy}\right|_{y=-0.4}$. And the Kolmogorov criterion for rapidity spectra of $\bar{\Lambda}$ for nuclear collisions at some energy:

$$F\left(\frac{dN(\bar{\Lambda})}{dy}, T\right) = \max \left| \left.\frac{dN^{\exp}(\bar{\Lambda})}{dy}\right|_{y_k} - \left.\frac{dN^{theory}(\bar{\Lambda})}{dy}\right|_{y_k} \right| = \left| \left.\frac{dN^{\exp}(\bar{\Lambda})}{dy}\right|_{y=-0.4} - \left.\frac{dN^{theory}(\bar{\Lambda})}{dy}\right|_{y=-0.4} \right|,$$

where *y* has all available indices *k* for all data points. The physical observable and used theory is indicated in brackets of *F*.

Having calculated of all $F(B_j, T_i)$ for all observables from the set $S_B$ and for all theories from the set $S_T$, we can estimate the adequacy of all theories in describing of the phenomenon *P*. To do this, we first make *F* independent of their units of measurement:

$$\chi_{worst}^2(B_j, T_i) = \left( \frac{\max \left| B_{j,k}^{\exp} - B_{j,k}(T_i) \right|}{\sigma^{\exp}(B_{j,k})} \right)^2, \qquad (2)$$

$$\delta_{worst}(B_j,T_i)=\left|\frac{\max\left|B_{j,k}^{\exp}-B_{j,k}(T_i)\right|}{B_{j,k}^{\exp}}\right|, \qquad (3)$$

that is, we included the Kolmogorov criterion (1) in the chi-square (2) and relative (3) criteria. According to (2 – 3), we use only one point of each observable $B_j^{\exp}$ from all its data points $n_{B_j}$ for analysis. This point has the maximum deviation from the corresponding point of the theoretically calculated observable $B_j(T_i)$. We called these criteria the worst, because with the maximum deviation of the compared distributions, the chi-square and the relative criterion are not necessarily maximum – the chi-square depends on the experimental error, and the relative criterion depends on the scale.

Suppose we have calculated $\chi^2_{worst}(B_j,T_i)$ and $\delta_{worst}(B_j,T_i)$ for all observables in the set $S_B$ and for all theories in the set $S_T$. Thus we have the set of chi-square and relative criteria. If the number of theories and observables is large, it will be very difficult to find the best theory, since, for example, in heavy ion collisions we have a huge number of observables and a dozen theories contradicting each other, which have different agreement with experiment depending on the energy of nuclear collisions, the centrality and from the analyzed physical observable. Therefore, we average all chi-square and relative criteria across all observables for each theory:

$$\left\langle \chi^2_{worst}(T_i)\right\rangle=\frac{1}{x}\sum_{j=1}^{x}\chi^2_{worst}(B_j,T_i), \qquad (4)$$

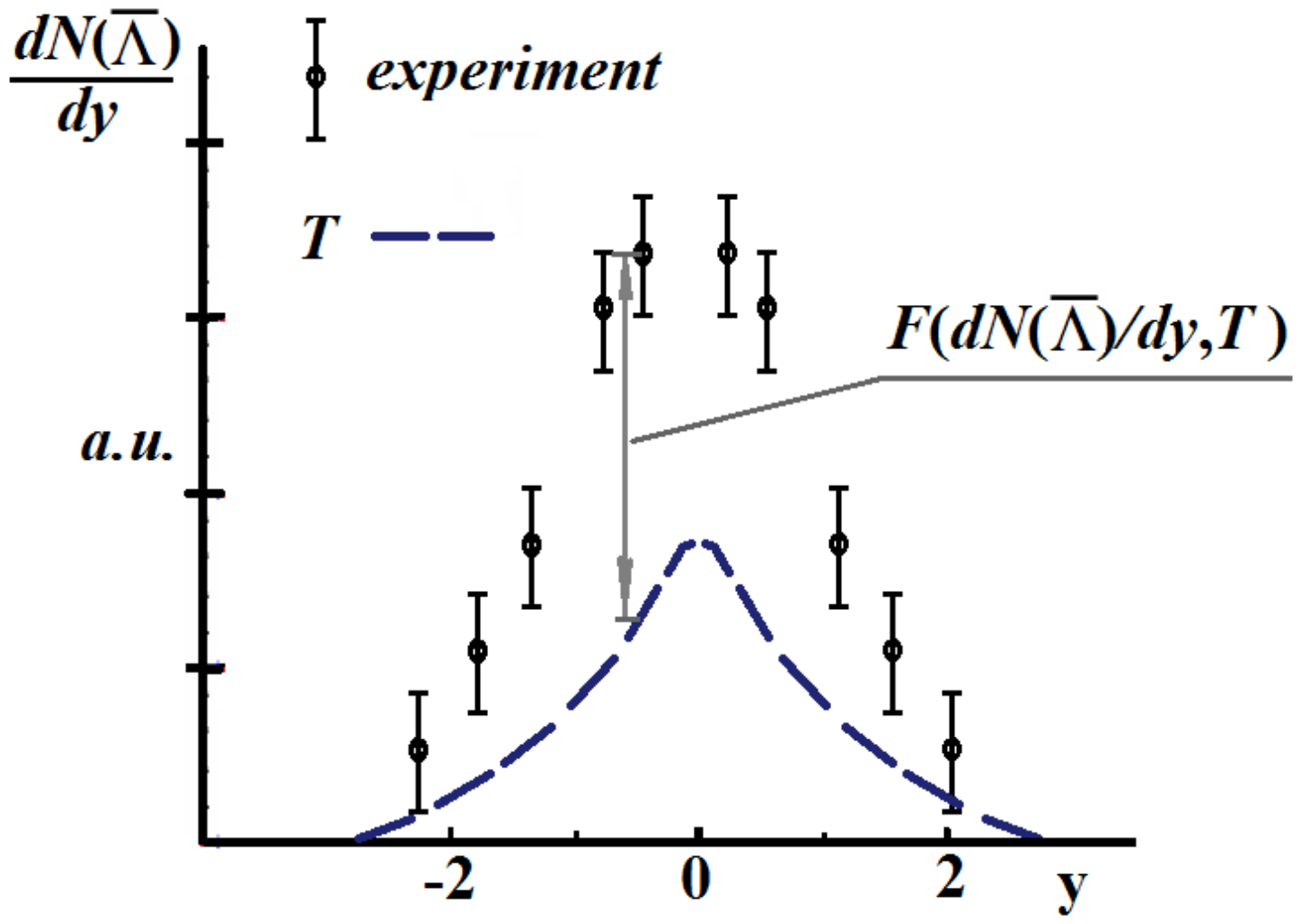

***Figure 1.*** *Rapidity spectra of $\bar{\Lambda}$ for ultrarelativistic nuclear collisions. Theoretical calculations based on some theory T are represented by the dashed line; experimental measurements are represented by circles with errors. At y = – 0.4, the deviation of the experimental point from the theory is maximal; this maximum deviation is the Kolmogorov criterion F.*

$$\left\langle \delta_{worst}(T_i)\right\rangle=\frac{1}{x}\sum_{j=1}^{x}\delta_{worst}(B_j,T_i), \qquad (5)$$

and we have a set of criteria of 2*k* elements, where k is the number of theories. In such way, we can find the best theory for which the chi-square (4) and relative (5) criteria are minimal compared to these criteria for other theories.

We calculated the errors as follows:

$$\sigma(\tilde{f})=\frac{1}{x}\cdot\sqrt{\sum_{j=1}^{x}\left[\left(\frac{\partial f}{\partial B_j^{\exp}}\cdot\sigma_j^{\exp}\right)^2+\left(\frac{\partial f}{\partial B_j^{th}}\cdot\sigma_j^{th}\right)^2\right]}, \qquad (5')$$

where $\tilde{f} = \langle \delta_{worst}(T_i) \rangle$ from (5) or $\langle \chi^2_{worst}(T_i) \rangle$ from (4), $f = \delta_{worst}(B_j, T_i)$ or $\chi^2_{worst}(B_j, T_i)$ respectively, $B_j^{\exp} \equiv B_{j,k}^{\exp}$ from (2) and (3), $B_j^{th} \equiv B_{j,k}(T_i)$ also from (2) and (3), and $\sigma_j^{\exp}$ is the experimental error in measuring of $B_{j,k}^{\exp}$, $\sigma_j^{th}$ is the theoretical error in calculating of $B_{j,k}(T_i)$. Formula (5′) is correct if $\sigma_j^{\exp,th} << B_j^{\exp,th}$, otherwise another formula was used to calculate the error:

$$\sigma(\tilde{f}) = \frac{1}{x} \cdot \sum_{j=1}^{x} \left[ c_j \cdot \sigma_j^{\exp} + b_j \cdot \sigma_j^{th} \right], \qquad (5'')$$

where the values of $c_j$ and $b_j$ depend on the type of functional dependence of $B_j^{\exp,th}$ in $\tilde{f}$.

We need to answer one question: is it correct that we exclude from the analysis all but one data point for each observable? Suppose some theory $T_i$ coincides with all points of the experimentally measured observable $B_j^{\exp}$ except for one point. If we include all points in the analysis as we did in [3], [4], then the points with good agreement between theory and experiment will suppress the one point with the worst agreement, which makes any criterion small. In this case, we mistakenly assume that the theory is good ignoring what appears to be just one bad point. But precisely this point corresponds to some physical processes within the phenomenon being studied, which are not included in the $T_i$ theory. A good theory should describe the phenomenon completely, not approximately. Therefore, we exclude all good data points from the analysis, leaving only one, which is the worst. For such an analysis, the Kolmogorov criterion (1) should be used.

### 3. Application of the method

The following set of theories was used: HSD, PHSD, HSDwCSR, PHSDwCSR. Hadron-String Dynamics without chiral symmetry restoration (HSD) transport approach, applied to experimental data for central nucleus-nucleus collisions from SIS/BEVALAC to upper RHIC energies, was taken from [8], [9], [10], [11], [12], [13], [14], [15], [16], [17], [18]. Parton-Hadron-String Dynamics without chiral symmetry restoration (PHSD) transport approach, applied to experimental data for central nucleus-nucleus collisions from upper SIS to LHC energies, was taken from [8], [9], [12], [13], [18], [19], [20], [21], [22]. Hadron-String Dynamics with chiral symmetry restoration (HSDwCSR) transport approach, applied to experimental data for central nucleus-nucleus collisions from AGS to SPS energies, was taken from [18]. Parton-Hadron-String Dynamics with chiral symmetry restoration (PHSDwCSR) transport approach – from [18]. PHSD differs from HSD by the inclusion of partonic degrees of freedom (QGP formation) in dynamic processes. HSD and PHSD have the addition in Figures 2 – 3 “w/o CRS” (without CRS).

The HSD and PHSD models have different versions, but all versions were combined according to the rule (6 – 7), since each subsequent version does not contradict the previous one, that is, the new included processes in the new version can be considered mutually compatible with the processes in the old versions:

$$\langle \chi^2_{worst}(HSD) \rangle = \frac{1}{n} \sum_{i=1.0}^{3.3} \langle \chi^2_{worst}(HSDi) \rangle, \qquad (6)$$

$$\langle \chi^2_{worst}(PHSD) \rangle = \frac{1}{n} \sum_{i=1.0}^{3.3} \langle \chi^2_{worst}(PHSDi) \rangle, \qquad (7)$$

where $i$ run from the first version (HSD1.0, PHSD1.0) to the version 3.3 (HSD3.3, PHSD3.3), $n$ is the number of versions. The meaning of formulas (6–7) is that for each version of the model we have own set of physical observables and we cannot use only the latest version for analysis, since the calculated observables in it will not be enough. The same procedure was done for relative criteria $\langle \delta_{worst}(HSD) \rangle$ and $\langle \delta_{worst}(PHSD) \rangle$. This was not done for the CSR models because we only used data for these models for the single versions from [18].

The analysis used the following set of observables, taken from the above-mentioned articles [8]–[22]: distribution of transverse mass $m_T$ or momentum $p_T$: $B_1 = \frac{1}{m_T}\frac{d^2N}{dm_T dy}(m_T)$, longitudinal rapidity $y$ distributions: $B_2 = \frac{dN}{dy}(y)$, hadron yields measured at midrapidity: $B_3 = \left.\frac{dN}{dy}\right|_{y=o}$, total yields measured within 4π solid angle: $B_4 = Y$, and dilepton invariant mass distributions: $B_5 = \frac{dN_{ll}}{dM_{ll}}(M_{ll})$. The first four observables were taken for light flavor (LF) and strange hadrons (S), and $B_1$ also for direct photons. The calculation of the worst relative criteria was carried out separately for each type of particles, and then averaged using the formulas in Figure 2. The same procedure was done for the worst chi-square criteria (Figure 3). For example, we calculate relative criteria for charged pions (in Figure 2 we used index $M_{LF}$ for them), compering experiment and the $i^{th}$ version of the HSD model:

$$\delta_{worst}(B_j, HSDi)_{M_{LF}} = \left|\frac{\max\left|B_{j,k}^{\exp} - B_{j,k}(HSDi)\right|}{B_{j,k}^{\exp}}\right|_{M_{LF}}, \quad (8)$$

Thus, we have 4 worst relative criteria for charged pions. Then we averaged them:

$$\left\langle\delta_{worst}(HSDi)\right\rangle_{M_{LF}} = \frac{1}{4}\sum_{j=1}^{4}\delta_{worst}(B_j, HSDi)_{M_{LF}}, \quad (9)$$

and as a result we have the worst relative criterion for charged pions for the $i^{th}$ version of the HSD model. Then we compute (9) for all HSD versions and average them as in (6):

$$\left\langle\delta_{worst}\right\rangle_{M_{LF}} = \frac{1}{n}\sum_{i=1.0}^{3.3}\left\langle\delta_{worst}(HSDi)\right\rangle_{M_{LF}}, \quad (10)$$

and as a result we have the worst criterion for charged pions for HSD phenomenology. If we have data separately for each pion charge, then a set of criteria (8) is calculated for each charge and similar criteria are averaged over the charge. For example, the criteria for the total yields of positive and negative pions are averaged, and this value is then used in (9).

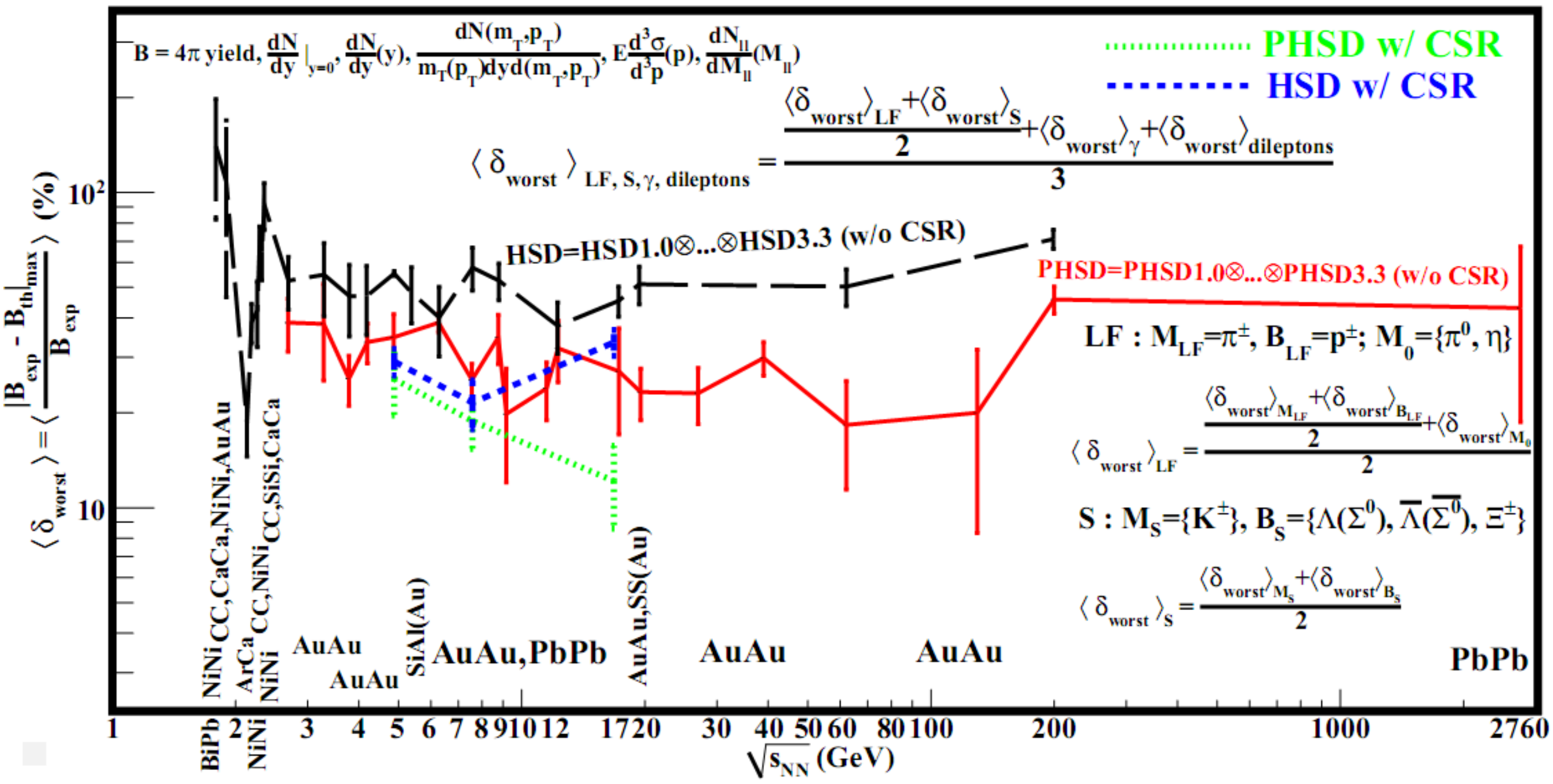

*Figure 2. The worst relative criteria of the agreement between models and experimental data as functions of the energy of central nuclear collisions. The formulas demonstrate the method described in the text. LF, S – sets for light flavor and strange hadrons,*

*respectively. The tensor product symbol shows that the multiplied versions of the models do not contradict each other. The points are connected by lines to guide the eye.*

The same procedure (8) – (10) was done for (anti-)protons ($B_{LF}$), neutral pions and η-mesons ($M_0$), for which we separately calculated $\langle \delta_{worst} \rangle_{LF}$ shown in Figure 2. There also shown the worst relative criteria for strange particles $\langle \delta_{worst} \rangle_{S}$ and the resulting worst criterion $\langle \delta_{worst} \rangle_{LF,S,\gamma,dileptons}$, where the symbol γ refers to direct photons. Figure 3 should contain the same formulas for the worst chi-square criteria, but we have omitted them for clarity of the picture. We also averaged the worst criteria for observables from different types of colliding nuclei within each model.

Taking into account the phenomenology from [18], [23] and [24], we provide the following interpretation of Figures 2 – 3 for central nuclear collisions. The separation of the $\langle \chi^2_{worst} \rangle$ criteria for the PHSD and HSD models in Figure 3 already at $\sqrt{s_{NN}}$ = 2.7 GeV (they only touch each other with errors) could be caused by QGP ignition at upper SIS energies (star at Figure 4). The relative criterion $\langle \delta_{worst} \rangle$ for PHSD in Figure 2 at this and higher nuclear collision energies lies systematically below the worst relative criterion for the HSD model. Since at upper SPS energies partonic matter contains no more than 40% of the nuclear collision energy [23], we can say that at upper SIS energies partonic matter should occupy a small part of the volume of the created fireball, and the rest of its volume consists of hadronic matter. That is, the fireball can be viewed as consisting of a drop of hot and dense exotic matter surrounded by a hadronic corona. The separation of the worst relative criteria for PHSD and HSD models at $\sqrt{s_{NN}}$ = 3.5 GeV can be explained by the transition of QGP phase to the Quarkyonic phase of matter [24] – the phase trajectory of a drop of exotic matter crosses the phase boundary (point 3.5 GeV in Figure 4).

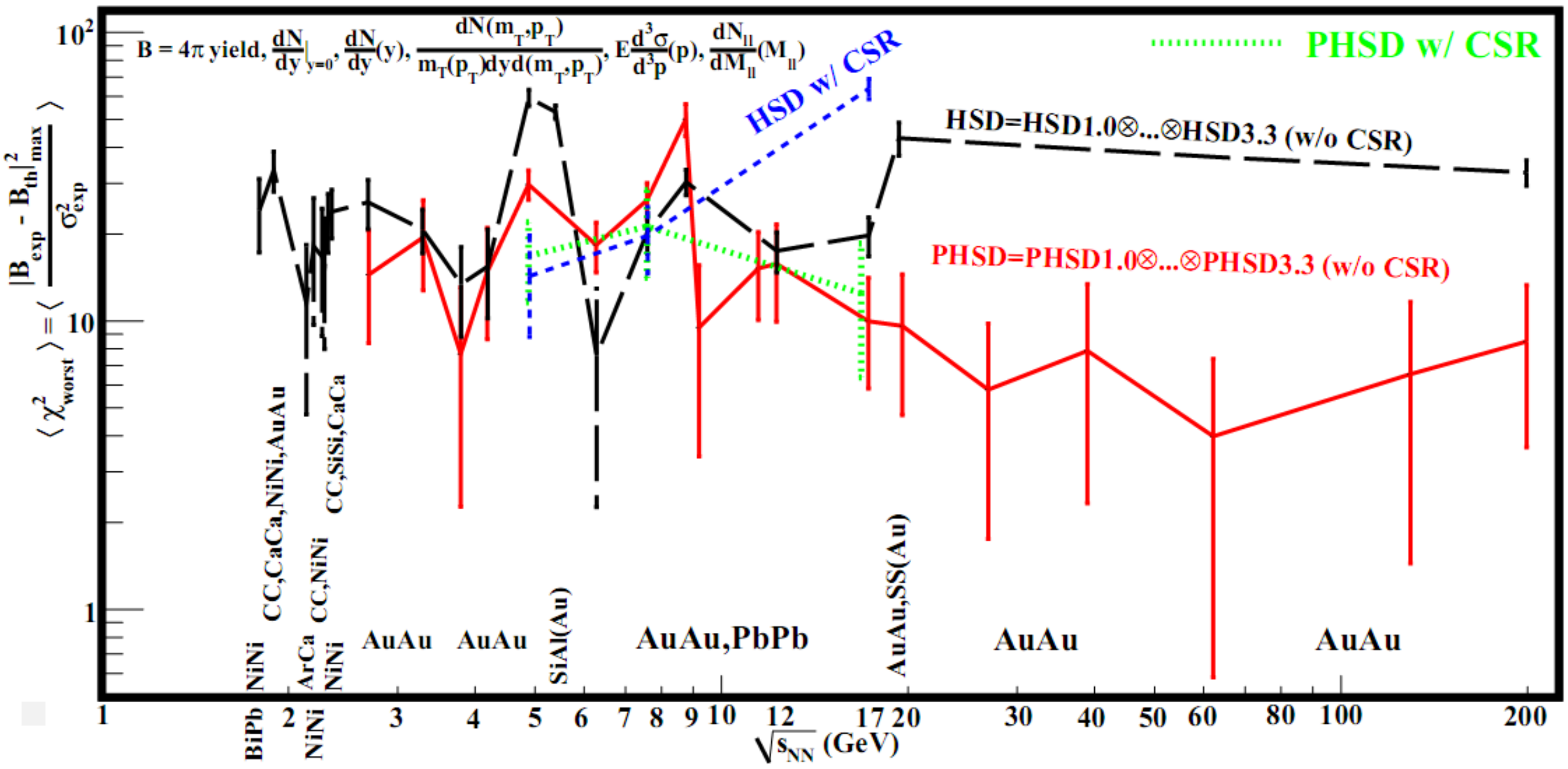

***Figure 3.*** *The worst chi-square criteria of the agreement between models and experimental data as functions of the energy of central nuclear collisions. The points are connected by lines to guide the eye.*

The phase trajectory of the hadronic corona remains in the hadronic phase at $\sqrt{s_{NN}}$ = 3.5 GeV – the Corona trajectory at the point 3.5 GeV in Figure 4. The energy pumped into the fireball by nuclear collisions is not enough to maintain the QGP phase in the droplet at $\sqrt{s_{NN}}$ = 3.5 GeV due to the larger volume of the fireball compared to its volume at lower energies (volume estimates are given below). The separations of the worst relative criteria in the interval $\sqrt{s_{NN}}$ = 4.3 ÷ 5.2 GeV (Figure 2) and $\langle \chi^2_{worst} \rangle$ criteria in the interval $\sqrt{s_{NN}}$ = 4.6 ÷ 5.5 GeV (Figure 3) for the PHSD and HSD models gives a hint that at energy of about $\sqrt{s_{NN}}$ = (4.3+4.6)/2 = 4.45 GeV, the phase trajectory of the droplet reaches the boundary between the Hadronic and Quarkyonic states of matter (point 4.4 GeV in Figure 4). That is, the energy pumped into the fireball at $\sqrt{s_{NN}}$ = 4.45 GeV is not enough to maintain the Quarkyonic phase of the droplet matter due to, again, the increased volume of the fireball.

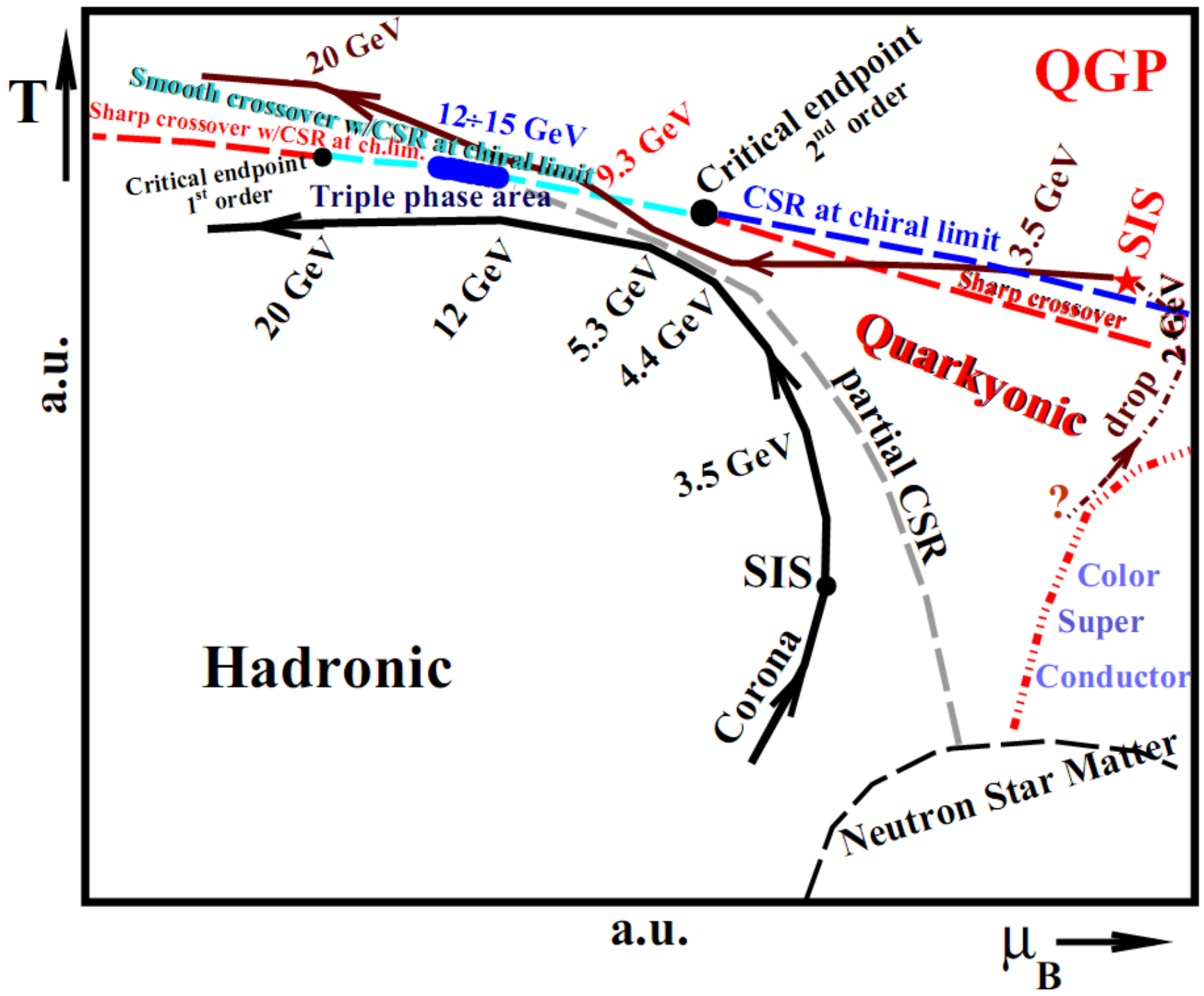

***Figure 4.*** *Phase diagram of strongly interacting matter produced in central nuclear collisions. The temperature T and baryon chemical potential $\mu_B$ are averaged over the entire space-time of the existence of a drop or corona (see text).*

There are two possibilities for the evolution of the drop, starting from $\sqrt{s_{NN}}$ = 4.45 GeV. Both of them are realized in the lab with their own probabilities due to the quantum nature of the fireball, as discussed in [4], [5]. The first possibility is that the phase trajectory of the drop crosses the boundary at $\sqrt{s_{NN}}$ = 4.4 GeV and the substance of the drop becomes hadronic. The phase trajectory of the drop remains in the Hadronic phase until the energy $\sqrt{s_{NN}}$ = (5.5+5.2)/2 = 5.35 GeV, where it returns to the Quarkyonic phase: we interpret the overturn of the $\langle \chi^2_{worst} \rangle$ criteria for HSD and PHSD relative to each other (Figure 3) and the coincidence of the $\langle \delta_{worst} \rangle$ criteria (Figure 2) at $\sqrt{s_{NN}}$ = 5.35 GeV as a return of the phase trajectory of the drop into the Quarkyonic phase after this energy. We have not depicted this possibility in Figure 4. The second possibility is that the phase trajectory of the drop does not cross the boundary at $\sqrt{s_{NN}}$ = 4.45 GeV, but goes along it (the matter remains in the Quarkyonic phase) to the point corresponding to the energy $\sqrt{s_{NN}}$ = 5.35 GeV (the point 5.3 GeV in Figure 4). That is, the energy pumped into the fireball is enough to return the phase trajectory of the drop matter to the Quarkyonic phase or push it far from the boundary.

A sharp decreasing of the $\langle \delta_{worst} \rangle$ and $\langle \chi^2_{worst} \rangle$ criteria for PHSD after $\sqrt{s_{NN}}$ = 8.8 GeV with their minima at $\sqrt{s_{NN}}$ = 9.2 GeV (Figures 2 – 3) and the discrepancy of the $\langle \delta_{worst} \rangle$ for PHSDwCRS and HSDwCRS after 9.2 GeV and their $\langle \chi^2_{worst} \rangle$ criteria after $\sqrt{s_{NN}}$ = 10 GeV (taking into account the criteria errors, Figures 2 – 3) we interpret as the existing of a boundary between the Quarkyonic and the QGP phase. Averaging $\sqrt{s_{NN}}$ at these points, we obtain the position of the boundary at: $\frac{8.8+9.2+9.2+10}{4}$ = 9.3 GeV. Since PHSD assumes a transition to QGP through a smooth (2nd order) crossover, we argue that at $\sqrt{s_{NN}}$ = 9.3 GeV there is a smooth crossover transition. Below we will show that at $\sqrt{s_{NN}}$ = 3.5 GeV there is a sharp (1st order) crossover transition. This means that at baryon chemical potentials higher than we have at $\sqrt{s_{NN}}$ = 9.3 GeV there is a critical endpoint of 2nd order. We may suspect that the critical endpoint of 2nd order cannot be achieved through nuclear collisions in labs [30], since the phase trajectory of the droplet goes far from it.

We interpret the similar behavior of two types of $\langle \delta_{worst} \rangle$ and $\langle \chi^2_{worst} \rangle$ (Figures 2 – 3) for the HSD and PHSD models with their intersections in around $\sqrt{s_{NN}}$ = 12.3 GeV as a subsequent movement of the phase trajectory of the drop, after $\sqrt{s_{NN}}$ = 9.3, along the transition boundary with a smooth crossover between the Quarkyonic and QGP phases up to the triple point. At a collision energy of $\sqrt{s_{NN}}$ = 12.3 GeV, the intersection of both types of criteria means equal agreement between the partonic and hadronic transport approaches with experimental data, which can be interpreted as the existence of a hadron-quark-gluon mixed phase [3]. If we take into account the errors of the criteria in Figures 2 – 3, the triple point region is extended from $\sqrt{s_{NN}}$ = 12 GeV to 15 GeV. Therefore, it is more convenient to call it a triple phase area (Figure 4). The phase trajectory of the droplet remains in the QGP phase after the triple phase area, since we see that after $\sqrt{s_{NN}}$ = 15 GeV the $\langle \delta_{worst} \rangle$ and $\langle \chi^2_{worst} \rangle$ (Figures 2 – 3) for PHSD models are always lower than for the HSD models.

The phase trajectory of the corona matter remains in the hadronic phase throughout the entire range of energies considered in central nuclear collisions. In Figure 4, a possible scenario for the phase trajectory of a droplet at low energies SIS/BEVALAC is shown with a dot-dashed line, but other models are needed to determine the features of nuclear matter.

We understand the temperature and baryon chemical potential in Figure 4 as values averaged over the entire space-time evolution of the droplet or corona. That is, let a drop (corona) exist during the time interval $\Delta_{d(c)} = \tau_f - \tau_0$, where $\tau_0$ is a time of formation of the drop (corona), and $\tau_f$ is the time of freezing-out, disappearing, of the drop (corona). The temperature of the drop is higher than the corona and increases towards the center of the drop. Moreover, it is a function of time. The average volume of a drop (corona) over the time interval of its existence is $\langle V_{d(c)} \rangle = \frac{1}{\Delta_{d(c)}} \int_{\tau_0}^{\tau_f} V_{d(c)}(t)\, dt$. Then for the average temperature of the drop (crown) we obtain: $\langle T_{d(c)} \rangle = \frac{1}{\Delta_{d(c)} \cdot \langle V_{d(c)} \rangle} \int_{\tau_0}^{\tau_f} \int_V T_{d(c)}(t,\vec{r})\, dt d^3\vec{r}$, $\langle T_d \rangle > \langle T_c \rangle$. We applied the same reasoning to the baryon chemical potential: $\langle \mu_{B_{d(c)}} \rangle = \frac{1}{\Delta_{d(c)} \cdot \langle V_{d(c)} \rangle} \int_{\tau_0}^{\tau_f} \int_V \mu_{B_{d(c)}}(t,\vec{r})\, dt d^3\vec{r}$.

Let us now estimate the radius and volume of the drop and the entire fireball formed during central collisions of nuclei. Let the experiment measure the yield of particles $Y^{exp}_{fball}$ created during the time interval of the fireball's existence. Let us assume that the phase of the droplet substance is *A*, and the phase of the corona substance is *B*. Let some theory $T_1$ describes the evolution of the fireball, not paying attention to the coexistence of phase *B* with phase *A*, assuming the existence of only phase *A*. Then theory $T_1$ predicts the yield of particles from phase *A*: $Y^{T_1}_A$. Now let theory $T_2$ describes the evolution of the fireball, ignoring the coexistence of phases *A* and *B*, assuming the existence of only phase *B*. Then theory $T_2$ predicts the yield of particles from phase *B*: $Y^{T_2}_B$. The experimental value $Y^{exp}_{fball}$ is equal to the sum of yields of particles from both phases: $Y^{exp}_{fball} = Y_A + Y_B$ and these summands cannot be experimentally measured separately. Let us assume that there are numbers (properties of created matter) $a_1$ and $a_2$ such that: $a_1 = Y^{exp}_{fball} / Y_A \approx \langle V_{fball} \rangle / \langle V_d \rangle$, $a_2 = Y^{exp}_{fball} / Y_B \approx \langle V_{fball} \rangle / \langle V_c \rangle$, where $V_{d,c,fball}$ are the average volumes of the drop, corona and fireball during their life time untill freeze-out. Then for the relative criterion we can write:

$$\delta = \left|\left(Y_{fball}^{\exp} - Y_{A(B)}^{T_{1(2)}}\right)/Y_{fball}^{\exp}\right| = \left|1 - Y_{A(B)}^{T_{1(2)}}/Y_{fball}^{\exp}\right| = \left|1 - Y_{A(B)}^{T_{1(2)}}/\left(a_{1(2)} \cdot Y_{A(B)}\right)\right|.$$

Let theories $T_1$ and $T_2$ be considered very good. That is, although they may give results different from the experiment, but at least approximately: $Y_{A(B)}^{T_{1(2)}} \approx Y_{A(B)}$. Therefore $\delta \approx \left|1-1/a_{1(2)}\right| \approx \left|1-\frac{V_{d(c)}}{V_{fball}}\right|$. Thus, the relative criterion averaged over yields of all types of particles allows us to estimate the average volume of the various phases of the fireball. More precisely, the averaged relative criterion for the yields of all types of particles shows the averaged volume not occupied by the phase, which is taken into account by the theory used. In [4], we used larger statistics for total and mid-rapidity yields than in this work. As a result, Figure 1 from [4] shows not only the agreement between theory and experiment, but also the relative volumes of Hadronic matter and QGP in the energy range from AGS to RHIC. For example, the average volume not occupied by Hadronic matter (see the curve for 3FD with the Hadronic EoS model [25] in Figure 1 of [4]) in a fireball created in central collisions of heavy ions at $\sqrt{s_{NN}}$ = 63 GeV is almost 100% (including criterion errors), which means that at this energy the QGP phase fills almost the entire volume of the fireball. At $\sqrt{s_{NN}}$ = 2.7 GeV, all three versions of the 3FD model (Hadronic and two QGP versions – with smooth crossover and first-order transitions [25]) in [4] have averaged relative criteria of around 15%. It is logical to assume that the average volume not occupied by Hadronic matter is 15%, that is, $V_{QGP}$ = 15%. If we take the radius of the fireball created at $\sqrt{s_{NN}}$ = 2.7 GeV to be 10 fm [26], then the radius of the QGP droplet will be about 5.3 fm. Likewise, the volume of the exotic droplet at $\sqrt{s_{NN}}$ = 3.2 GeV (Figure1 of [4]) is about 10% of the total fireball volume, so the droplet radius is about 5.1 fm if the fireball radius has been increased to 11 fm. Let us assume a QGP density of 0.7 fm$^{-3}$ [26], then the average distance between partons will be 0.55 fm. Accepting reasoning that if the germ of a new phase, immersed in another phase, has a size (5.1 fm) between the average interparticle distance (0.55 fm) and the system size (11 fm), then we conclude, that this is a mesoscopic system "with deconfinement being rather a sharp crossover" [27] (i.e., 1$^{st}$ order crossover). So, at collision energies around $\sqrt{s_{NN}}$ = 3.5 GeV in Figure 4 the preferred transition is the sharp crossover, and taking into account the reasoning from [28], this sharp crossover split off from the chiral symmetry restoration (CSR) transition. Their boundaries meet at the critical endpoint at $\sqrt{s_{NN}}$ = 9.3 GeV, or below it, and above 5.3 GeV, as we discussed earlier. According to [28], this splitting is small, and the temperature of CSR is higher than the deconfinement temperature (in our case, sharp crossover).

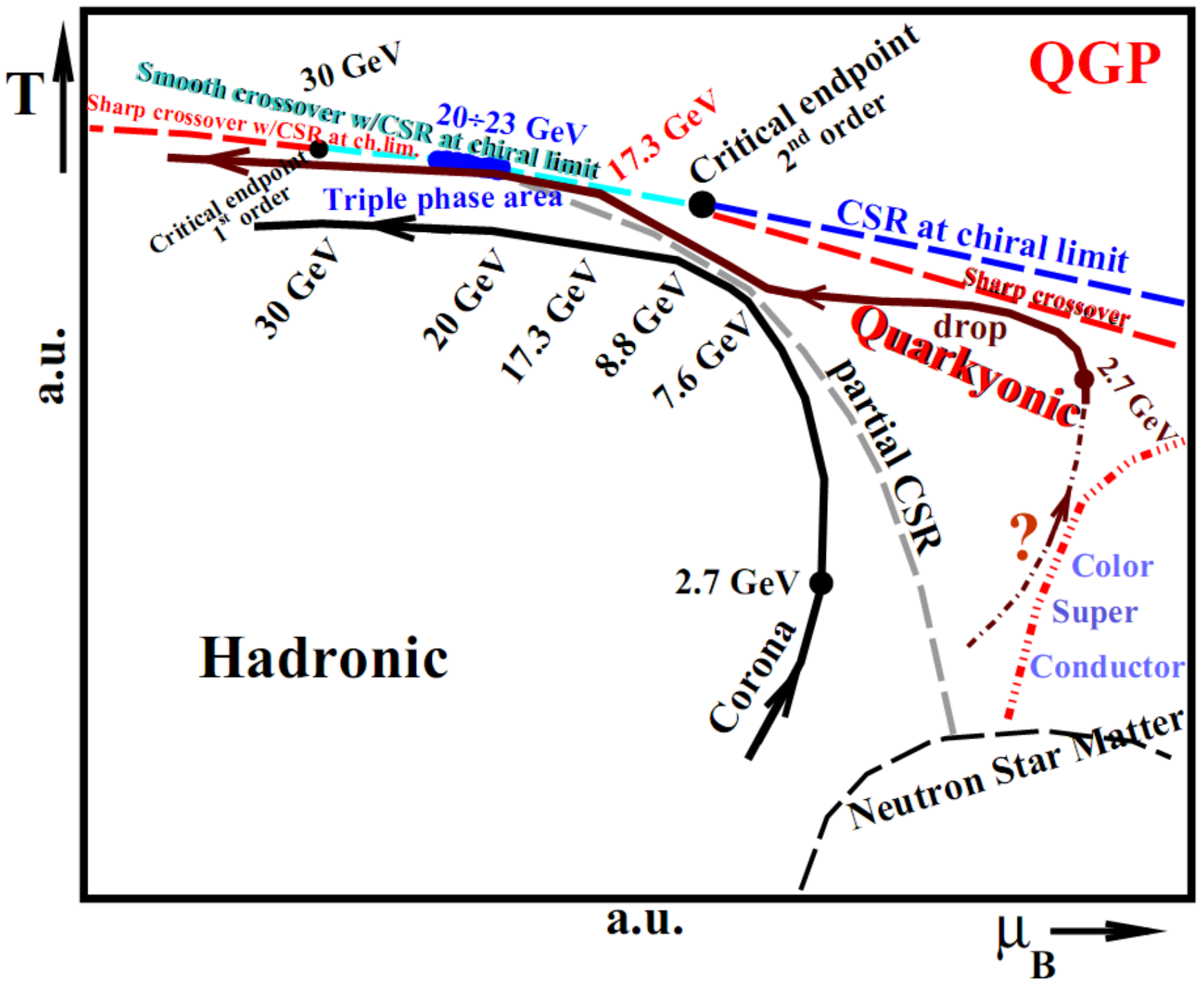

***Figure 5.*** *Phase diagram of strongly interacting matter produced in the mid-central heavy ion collisions. The temperature T and baryon chemical potential $\mu_B$ are averaged over the entire space-time of the existence of a drop or corona.*

We interpret the minima of both types of worst criteria for HSD at around $\sqrt{s_{NN}}$ = 2 GeV (Figures 2 – 3) as a deconfinement transition to the QGP state (Figure 4, dot-dashed line for the droplet). Figures 2 – 3 show that at $\sqrt{s_{NN}}$ = 4.9 GeV both types of the worst criteria for the HSDwCSR and PHSDwCSR models are smaller than those for the HSD and PHSD, what allows us to conclude that chiral symmetry is partially restored at around $\sqrt{s_{NN}}$ = 4.9 GeV. This means that chiral order parameter (with respect to the quark mass) decreases, starting from the boundary of the transition of Hadronic matter to the Quarkyonic. The chiral order parameter becomes zero (chiral limit) at the boundary of the transition of Quarkyonic matter to QGP at $\sqrt{s_{NN}}$ = 2 GeV and $\sqrt{s_{NN}}$ = 3.5 GeV (blue dashed line in Figure 4).

The smaller values of $\langle \delta_{worst} \rangle$ for PHSDwCSR compared to PHSD w/o CSR between $\sqrt{s_{NN}}$ = 15 and 17 GeV (Figure 2) indicate that a smooth crossover with CSR in chiral limit between the Hadronic matter and QGP occurs in this energy interval (Figure 4). We extend this boundary to $\sqrt{s_{NN}}$ = 20 GeV. $\langle \chi^2_{worst} \rangle$ coincide for both models at these energies, Figure 3. We interpret the kink of both types of the worst criteria for HSD and PHSD before and after $\sqrt{s_{NN}}$ = 20 GeV (Figures 2 – 3) as the existence of some singularity around this energy. Based on [29], at high temperature and low baryochemical potential, there is a rapid (sharp) crossover. Therefore, we assume a sharp crossover with CSR in chiral limit after $\sqrt{s_{NN}}$ = 20 GeV (Figure 4, red dashed line). That is, we have a critical endpoint of first order at around $\sqrt{s_{NN}}$ = 20 GeV.

By analyzing the Figure 2 from [4] and Figures 4 of this work, we have constructed the phase diagram of strongly interacting matter created in mid-central collisions of heavy ions at energies $\sqrt{s_{NN}}$ = 2.7 ÷ 27 GeV, depicted in Figure 5. We do not assume that the phase trajectory of the droplet reaches the deconfinement boundary, since both types of criteria for 3FD crossover EOS have very large values starting from $\sqrt{s_{NN}}$ = 3.2 GeV. That is, at energies $\sqrt{s_{NN}}$ = 3.2 GeV and above, the energy pumped into the fireball is not enough to ignite the QGP state in mid-central collisions. The positions of singularities of the nuclear matter are shifted towards higher energies of the mid-central heavy-ion collisions (Figure 5) compared to the central ones.

Recent experiments [31], [32], [33] and theoretical calculations [34] have shown that at heavy ions collision energies in the region around $\sqrt{s_{NN}}$ = 3 GeV, hadronic interactions dominate, while the contribution of partonic interactions is unclear - are they either completely absent or their contribution is insignificant? A theoretical calculation based on the 3FD model [34] with a transition to the partonic phase agrees with experiment no worse than a similar calculation using the same model without a transition to the partonic phase for the transverse-momentum and rapidity distributions of protons and light nuclei. In calculations using the 3FD model with the crossover transition from the hadronic to the QGP phase, it was shown that this transition should occur at the collision energy of heavy ions (Au + Au) between $\sqrt{s_{NN}}$ = 3 and 4.5 GeV [35].

We considered only scenarios with crossover transitions, limited by the phenomenology of the theoretical models used. In [5], it was shown that the fireball created in relativistic nuclear collisions has a quantum nature. Therefore, the evolution of a droplet can follow a different scenario, according to phenomenology not considered here. Thus, in [4] it was concluded that a smooth crossover and a first-order phase transition of Hadronic matter into the QGP state can occur with the same probabilities, for example, at energies of central collisions of heavy ions above $\sqrt{s_{NN}}$ = 12 GeV. At energies of mid-central heavy ion collisions in around $\sqrt{s_{NN}}$ = 2.7 GeV, we do not exclude that the phase trajectory of the drop reaches the QGP phase – this possibility is not depicted in Figure 5. The quantum state of a fireball created in ultrarelativistic nuclear collisions is a superposition of different states, what is shown in Figure 6.

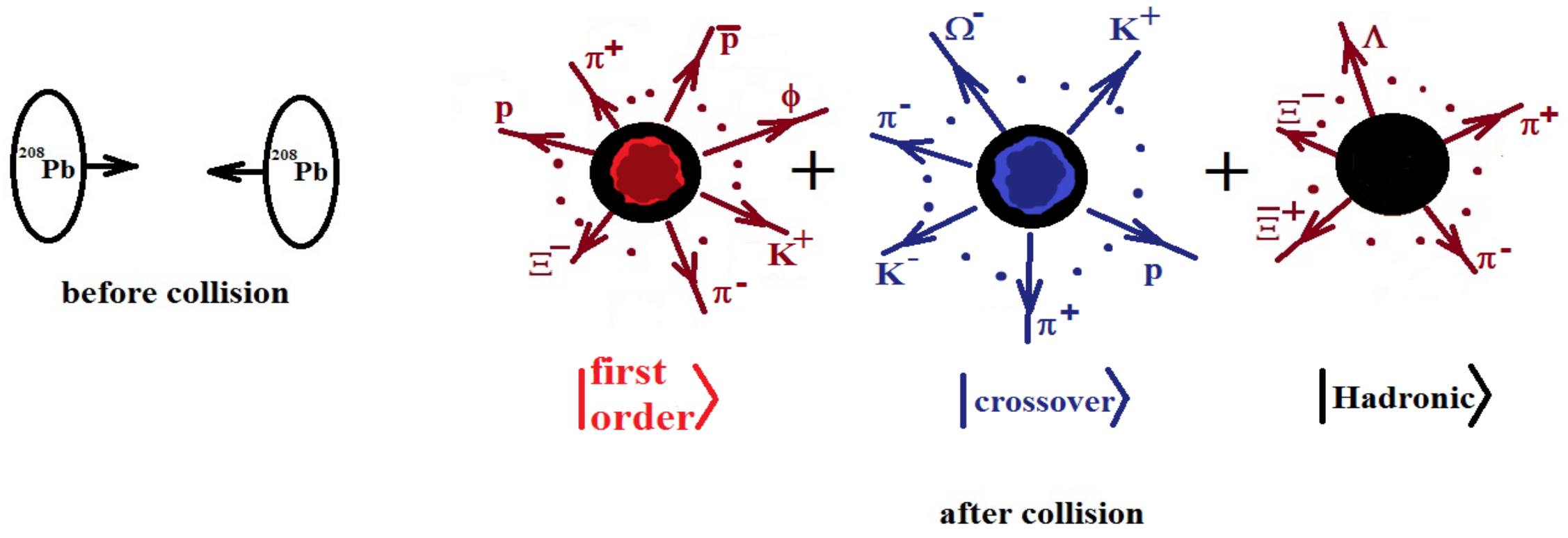

***Figure 6.*** *Diagram of superposition of different quantum states of a fireball created at SPS energies in ultra-relativistic nuclear collisions. State vectors are explained in the text.*

At diagram of Figure 6, the different quantum states of the fireball, for example, at SPS energies are realized with different probabilities. State vector |first order⟩ represents the evolution of a fireball through a first order phase transition of the droplet matter to the QGP state, |crossover⟩ – through a smooth or sharp crossover depending on the collision energy, |Hadronic⟩ – absence of ignition of the exotic phase in the droplet. The black circles in the two states of the diagram in Figure 6 represent a Hadronic corona enveloping a droplet of exotic matter. The evaporation of particles by a fireball is shown by lines with arrows, and the dots are the lines of other particles.

## 4. Conclusion

The use of meta-analysis and Kolmogorov criteria made it possible to separate the HSD and PHSD models already at energy $\sqrt{s_{NN}}$ = 2.7 GeV of central nucleus-nucleus collisions. The mapping corresponding phenomenology on plots of the Kolmogorov criteria versus the energy of nuclear collisions made it possible to clarify the possible positions of the critical endpoint of second order. This point has the baryon chemical potential larger than we have at central nuclear collisions at $\sqrt{s_{NN}}$ = 9.3 GeV and with lower temperature, which cannot be reached by the phase trajectory of the droplet. The critical endpoint of first order is localized at $\sqrt{s_{NN}}$ = 20 GeV, the triple phase area occupy the interval $\sqrt{s_{NN}}$ = 12 ÷ 15 GeV. At collision energy $\sqrt{s_{NN}}$ = 3.5 GeV, we have a splitting of the deconfinement transition into two ones: a sharp crossover (1$^{st}$ order crossover) transition and CSR in chiral limit (the latter at a higher temperature) between Quarkyonic matter and the QGP. The boundary of the transition with partial CSR between Hadronic and Quarkyonic matter is localized in the range $\sqrt{s_{NN}}$ = 4.4 ÷ 5.3 GeV, although the phase trajectory of a drop of hot matter does not cross it, that is, at lower energies of central collisions the phase trajectory of the drop is in the Quarkyonic phase. The boundary of a smooth crossover (2$^{nd}$ order crossover) transition with CSR in chiral limit between Quarkyonic matter and QGP is localized in the interval of central nuclear collisions $\sqrt{s_{NN}}$ = 9.3 ÷ 12 GeV, and between Hadronic matter and QGP in the interval $\sqrt{s_{NN}}$ = 15 ÷ 20 GeV. After $\sqrt{s_{NN}}$ = 20 GeV we have the boundary of a sharp crossover transition with CSR in chiral limit between Hadronic matter and QGP.

Ignition of a QGP droplet occurs when the phase trajectory of the droplet passes through a split transition of sharp crossover and CSR in chiral limit at around $\sqrt{s_{NN}}$ = 2 GeV. The volume of this drop occupies about 15% of the total volume of the fireball.

Taking into account one of the principles of quantum mechanics that if the existence of a certain state of a quantum system does not has internal contradictions, then this state must necessarily be realized in nature (laboratory). Then if in competing theories, in each separately, there is also no internal contradictions, then these theories are correct. Thereby, in central nuclear collisions at any energy (from $\sqrt{s_{NN}}$ = 2 GeV to $\sqrt{s_{NN}}$ = 2.76 TeV) there is a nonzero probability of non-ignition of the QGP, that is, there are events without QGP ignition. It is currently impossible to separate different scenarios for the evolution of a fireball experimentally, since it is necessary to carry out this separation in each nuclear collision, which is not yet technically feasible.

## References

[1] Laplace P-S. Théorie Analytique des Probabilités. Oeuvres Complètes **7** (3rd edition). Paris: Courcier 1820: lxxvii.

[2] Airy G. B. On the Algebraical and Numerical Theory of Errors of Observations and the Combination of Observations. London: Macmillan and Company, 1861.

[3] V. A. Kizka, V. S. Trubnikov, K. A. Bugaev, D. R. Oliinychenko. A possible evidence of the hadron-quark-gluon mixed phase formation in nuclear collisions. arXiv:1504.06483 [hep-ph]. https://doi.org/10.48550/arXiv.1504.06483

[4] V. Kizka. Comparison of the 3-Fluid Dynamic Model with Experimental Data. Nuclear Science. 2023, 7(3), 39-44. https://doi.org/10.11648/j.ns.20220703.11

[5] V. A. Kizka. On the Quantum Nature of a Fireball Created in Ultrarelativistic Nuclear Collisions. New Frontiers in Physical Science Research. 2022, 1, 52–62. https://doi.org/10.9734/bpi/nfpsr/v1/3541A

[6] The STAR Collaboration. Production of Protons and Light Nuclei in Au+Au Collisions at $\sqrt{s_{NN}}$ = 3 GeV with the STAR Detector. arXiv:2311.11020 [nucl-ex]. https://doi.org/10.48550/arXiv.2311.11020

[7] W. Feller. On the Kolmogorov-Smirnov limit theorems for empirical distributions. Annals of Math. Stat. 1948, 19, 177–189. https://doi.org/10.1214/aoms/1177730243

[8] W. Cassing et al. Parton/hadron dynamics in heavy-ion collisions at FAIR energies. EPJ Web of Conferences. 2015, 95, 01004. https://doi.org/10.1051/epjconf/20159501004

[9] W. Cassing, E.L. Bratkovskaya. Parton-Hadron-String Dynamics: an off-shell transport approach for relativistic energies. Nucl. Phys. A. 2009, 831, 215-242. https://doi.org/10.1016/j.nuclphysa.2009.09.007

[10] J. Geiss, W. Cassing, C. Greiner. Strangeness Production in the HSD Transport Approach from SIS to SPS energies. Nucl. Phys. A. 1998, 644, 107–138. https://doi.org/10.1016/S0375-9474(98)80011-0

[11] W. Cassing, E. L. Bratkovskaya. Hadronic and electromagnetic probes of hot and dense nuclear matter. Physics Reports. 1999, 308, 65–233. https://doi.org/10.1016/S0370-1573(98)00028-3

[12] E. L. Bratkovskaya et al. The QGP phase in relativistic heavy-ion collisions. Proceedings of the International Symposium on Exciting Physics. Makutsi-Range, South Africa, 2011, 13 – 20 November. https://doi.org/10.1007/978-3-319-00047-3_19

[13] E. L. Bratkovskaya et al. Properties of the partonic phase at RHIC within PHSD. 27th Winter Workshop on Nuclear Dynamics (WWND) in Colorado. 2011, USA on February 6 – 13. https://doi.org/10.1088/1742-6596/316/1/012027

[14] T. Anticic et al. (NA49 Collaboration). System-size and centrality dependence of charged kaon and pion production in nucleus-nucleus collisions at 40 AGeV and 158 AGeV beam energy. Phys. Rev. C. 2012, 86, 054903. https://doi.org/10.1103/PhysRevC.86.054903

[15] T. Anticic et al. (NA49 Collaboration). System-size dependence of Lambda and Xi production in nucleus-nucleus collisions at 40A and 158 AGeV measured at the CERN Super Proton Synchrotron. Phys. Rev. C. 2009, 80, 034906. https://doi.org/10.1103/PhysRevC.80.034906

[16] E. L. Bratkovskaya et al. Strangeness dynamics and transverse pressure in relativistic nucleus-nucleus collisions. Phys. Rev. C. 2004, 69, 054907. https://doi.org/10.1103/PhysRevC.69.054907

[17] H. Weber, E. L. Bratkovskaya, W. Cassing, H. Stoecker. Hadronic observables from SIS to SPS energies - anything strange with strangeness? Phys. Rev. C. 2003, 67, 014904. https://doi.org/10.1103/PhysRevC.67.014904

[18] W. Cassing, A. Palmese, P. Moreau, E. L. Bratkovskaya. Chiral symmetry restoration versus deconfinement in heavy-ion collisions at high baryon density. Phys. Rev. C. 2016, 93, 014902. https://doi.org/10.1103/PhysRevC.93.014902

[19] V. P. Konchakovski, W. Cassing, V. D. Toneev. Impact of the initial size of spatial fluctuations on the collective flow in Pb-Pb collisions at $\sqrt{s_{NN}}$ = 2.76 TeV. J. Phys. G: Nucl. Part. Phys. 2015, 42, 055106. https://doi.org/10.1088/0954-3899/42/5/055106

[20] O. Linnyk et al. Hadronic and partonic sources of direct photons in relativistic heavy-ion collisions. Phys. Rev. C. 2015, 92, 054914. https://doi.org/10.1103/PhysRevC.92.054914

[21] Pierre Moreau et al. (Anti-)strangeness production in heavy-ion collisions. Proceedings of the 15th International Conference on Strangeness in Quark Matter (SQM2015). 2015, 6-11 July, JINR, Dubna, Russia. https://doi.org/10.1088/1742-6596/668/1/012072

[22] E. L. Bratkovskaya et al. Dilepton production from SIS to LHC energies. Proceedings of the 28th Winter Workshop on Nuclear Dynamics. 2012, Dorado del Mar, Puerto Rico, April 7-14. https://doi.org/10.1088/1742-6596/389/1/012016

[23] O. Linnyk, E. L. Bratkovskaya, W. Cassing. Strangeness production within Parton-Hadron-String Dynamics (PHSD). J. Phys. G: Nucl. Part. Phys. 2010, 37, 094039. https://doi.org/10.1088/0954-3899/37/9/094039

[24] A. Andronic et al. Hadron Production in Ultra-relativistic Nuclear Collisions: Quarkyonic Matter and a Triple Point in the Phase Diagram of QCD. Nucl. Phys. A. 2010, 837, 65–86. https://doi.org/10.1016/j.nuclphysa.2010.02.005

[25] Ivanov Yu. B. Alternative Scenarios of Relativistic Heavy-Ion Collisions: I. Baryon Stopping. Phys. Rev. C. 2013, 87, 064904. https://doi.org/10.1103/PhysRevC.87.064904

[26] D. A. Fogaca, S. M. Sanches Jr., R. Fariello, F. S. Navarra. Bubble dynamics and the quark-hadron phase transition in nuclear collisions. Phys. Rev. C. 2016, 93, 055204. https://doi.org/10.1103/PhysRevC.93.055204; arXiv:1601.04596v1 [hep-ph].

[27] V.I. Yukalov, E.P. Yukalova. Models of mixed hadron-quark-gluon matter. Proc. Sci. (ISHEPP). 2012, 046. https://doi.org/10.48550/arXiv.1301.6910

[28] C. A. Dominguez, M. Loewe, Y. Zhang. Chiral symmetry restoration and deconfinement in QCD at finite temperature. Phys. Rev. D. 2012, 86, 034030. https://doi.org/10.1103/PhysRevD.86.034030

[29] Y. Aoki et al. The order of the quantum chromodynamics transition predicted by the standard model of particle physics. Nature. 2006, 443, 675–67. https://doi.org/10.1038/nature05120

[30] The NA61/SHINE Collaboration. Search for a critical point of strongly-interacting matter in central $^{40}$Ar +$^{45}$Sc collisions at 13A-75A GeV/c beam momentum. arXiv:2401.03445 [nucl-ex]. https://doi.org/10.48550/arXiv.2401.03445

[31] Barnabas Porfy. Femtoscopy at NA61/SHINE using symmetric Lévy sources in central 40Ar+45Sc from 40A GeV/c to 150A GeV/c. arXiv: 2406.02242 [nucl-ex]. https://doi.org/10.48550/arXiv.2406.02242

[32] Dylan Neff. Probing the nature of the QCD phase transition with higher-order net-proton number fluctuation and local parton density fluctuation measurements at RHIC-STAR. arXiv: 2405.20929 [nucl-ex]. https://doi.org/10.48550/arXiv.2405.20929

[33] Dylan Neff. Recent Highlights from STAR BES Phase II. arXiv: 2405.20928 [nucl-ex]. https://doi.org/10.48550/arXiv.2405.20928

[34] M. Kozhevnikova, Yu. B. Ivanov. Light Nuclei Production in Au+Au Collisions at √sNN = 3 GeV within Thermodynamical Approach: Bulk Properties and Collective Flow. arXiv: 2311.08092 [nucl-th]. https://doi.org/10.48850/arXiv.2311.08092

[35] Yu. B. Ivanov, M. Kozhevnikova. What Can We Learn from Directed Flow at STAR-FXT Energies? arXiv: 2403.02787 [nucl-th]. https://doi.org/10.48850/arXiv.2403.02787